\documentstyle[12pt]{article}   
\begin{document}   
\thispagestyle{empty}   
\rightline{UOSTP-02105}  
\rightline{{\tt hep-th/0208046}}   
   
\

\   
   
\vskip 0cm   
\centerline{  
\Large   
\bf Thoughts on  Big Bang   
}   
 
\vskip .2cm    
 
\vskip 1.4cm 
\centerline{  
Dongsu Bak
} 
\vskip 10mm  
\centerline{ \it 
Physics Department,  
University of Seoul, Seoul 130-743, Korea}
\centerline{ \it
e-mail: dsbak@mach.uos.ac.kr}  
\vskip 3mm

\vskip 1.6cm    
\begin{quote}    
{We present a consistent framework that enables us to understand the
big bang singularity of our
universe.
}
\end{quote}   
   

\newpage   
\baselineskip 17pt   

Until now there has been no theoretical framework to explain 
the big bang singularity consistently. String theories 
in principle
include consistent gravity theories but so far there is no way
to address  big-bang-like singularities in a controlled manner.

In this short note, we like to attempt to present a consistent way
of addressing  such cosmological big-bang-like singularities. 

One of the key observations lies in the fact  that 
our universe should start out as a low entropic state and its entropy
grows afterward according to
the generalized second law.
In fact the present date entropy including gravitational one may be 
estimated around $10^{120}$ and the entropy near big bang
is considerably smaller and almost nothing compared to the 
present value. Explaining how such low entropic states arise 
naturally is a question seemingly impossible to answer.
This is because the low entropic state immediately implies
that it should be an improbable state at any rate. 
Thus we are facing  the problem of explaining the arising of 
such improbable 
states that cannot be natural in any theoretical frameworks.
Inflationary scenario is a good phenomenological attempt for the choice 
of the initial states which may lead to the present state of 
universe but cannot explain why our universe begins with such  low
entropic states after all.

In this note, we like to address the low entropy problem in a totally 
different manner. We assume that the gravitational theories
governing the dynamics of our universe has a dual description
in terms of quantum mechanical system of finite number of
degrees. This assumption is based on the recent development of
AdS/CFT correspondence\cite{maldacena} and 
understanding of holographic principle\cite{thooft}.
Further, 
recent observational data indicates that there presents a nonvanishing
positive vacuum energy associated with our universe\cite{perlmutter}. 
If this 
is identified with cosmological constant, which by definition
cannot  be relaxed to zero by any classical means, one  has to deal 
with a spacetime with positive cosmological constant.
In this case, one may argue the
 degrees of freedom  in the universe with positive cosmological
constant is limited by ${ 3 /( 8 G^2 \Lambda})$ with $\Lambda$
being the cosmological constant\cite{bousso}.

The time evolution of such finite quantum mechanical system but having
enough number of degrees has the 
following characters.
Along infinite interval of time i.e. $-\infty <t < + \infty$, 
the system mostly remains in its
equilibrium states which correspond to the maximally entropic states
for a given set of macroscopic conserved quantities. 
Of course, there are always 
fluctuations of all kinds of  sizes around the equilibrium states of 
the quantum mechanical system. The larger the fluctuations are, 
one has in general the smaller chances of occurrence.  

We like to identify the big bang and subsequent cosmic time evolution
 as a really huge fluctuation
above the equilibrium  occurring 
rarely along the evolution of the system. 
At the peak of one such fluctuation,
the state displaced farthest from the equilibrium occurs along the
one cycle of rising and subsiding of the fluctuation. 
If the fluctuation is huge enough,
the peak corresponds 
to a considerably low entropic state compared to the equilibrium states.
 This peak will be identified 
 with the big bang event and, thus, one has 
a natural explanation of the low entropy nature of big bang.

At the peak a violent relaxation toward equilibrium will follow, which 
certainly matches with the violent nature of big bang of our universe.
For the huge enough fluctuation, the relaxation will occur for a cosmic 
time scale, which corresponds to the cosmological  evolution
we observe now. After the peak  the entropy of the system 
tends to grow in accordance
with the second law. Before the peak, however, the entropy is getting
smaller for a cosmic time scale as the rising of the fluctuation.
This is not in a contradiction with the second law because we are talking
about  fluctuations of rare occurrences.

Of course, the above framework does not provide any reason 
why the present state
of our specific universe  
among all kind of possibilities is actually resulted from the big bang event.
We are  certainly not attempting such explanations.
Here we like to merely point out that the above framework may set up
a consistent arena in addressing big-bang-like singularities.
Such attempts, anyway, will involve the argument of
anthropic principle. For example, the universe of the above framework 
will remain at equilibrium
states most of the time, which correspond to  non-livable states
and may be disregarded from the view point of the 
anthropic principle.   

Finally, it should be commented that, in the above discussion, 
we implicitly identify the time in the quantum mechanical system
with that in the gravity theory of 
our universe. This may be understood as follows. Due to the generalized
second law holding both of the systems (descriptions), the directions
of the time should be identified. Further utilizing the general covariance
of the gravity theory, the time in the gravity theory can be identified
with that in the quantum mechanical system.

We are eagerly looking for observational evidences supporting
the above framework and further studies are required in this direction.

{\sl Note added: We are informed that a similar scenario is presented
in Ref.~\cite{susskind}.}


\noindent{\large\bf Acknowledgment} We would like to thank Jihn E. Kim
for the valuable comments. This work is supported in part by
KOSEF 1998 Interdisciplinary Research Grant 
and by UOS 2002 Academic Research Grant.
 


\end{document}